\documentclass[eqsecnum,prd,aps,showpacs]{revtex4}
\newcommand{\beq}{\begin{equation}}
\newcommand{\eeq}{\end{equation}}
\newcommand{\bqa}{\begin{eqnarray}}
\newcommand{\eqa}{\end{eqnarray}}
\newcommand{\fr}{\frac}
\begin{document}
\def\g{\beta}
\title{The spectrum of endstates of gravitational collapse with tangential stresses}
\author{S\' ergio M. C. V. Gon\c calves}
\altaffiliation{Current address: Department of Physics, Yale University, New Haven, CT 06511}
\affiliation{Theoretical Astrophysics, California Institute of Technology, Pasadena, California 91125}
\author{Sanjay Jhingan}
\affiliation{Yukawa Institute for Theoretical Physics, Kyoto University, Kyoto 606-8502, Japan}
\author{Giulio Magli}
\affiliation{Dipartimento di Matematica, Politecnico di Milano, Piazza Leonardo da Vinci, 32, 20133 Milano, Italy}
\date{July 26, 2001}
\preprint{DRAFT $\#1$}
\begin{abstract}
The final state---black hole or naked singularity---of the gravitational collapse of a marginally bound matter configuration in the presence of tangential stresses is classified, in full generality, in terms of the initial data and equation of state. If the tangential pressure is sufficiently strong, configurations that would otherwise evolve to a spacelike singularity, result in a locally naked singularity, both in the homogeneous and in the general, inhomogeneous density case.
\end{abstract}
\maketitle

\section{Introduction} 

The final state of gravitational collapse remains as one of the 
outstanding problems in classical general relativity. For a given 
set of Cauchy data, hyperbolic evolution of the field 
equations leads to a unique final state and, under certain conditions,
a singularity is formed~\cite{hawking&ellis73}. However, it is not known whether
this singularity is spacelike, or a visible spacetime singularity. 
Understanding how such singularities arise from regular initial 
data, and whether they can be visible (at least to local 
observers), is still to a large extent an open issue~\cite{r}.

The most studied analytical solution is that describing 
spherical dust collapse, whose detailed analyses over 
the past ten years have provided many valuable insights into the 
formation, visibility, and causal structure of 
singularities~\cite{review}. Albeit of considerable interest in 
its own right, dust collapse constitutes a highly idealized model, 
both in terms of geometry and matter content. Whilst the former 
may arguably be a reasonable approximation for astrophysical 
collapse~\cite{nakamura82}, the existence of pressures in high 
density regimes, together with an effective equation of state, 
cannot be neglected in realistic situations~\cite{miller&sciama79}. 
It would, therefore, be highly desirable
to understand the final state of spherical gravitational collapse for
generic equations of state. Unfortunately, this objective 
looks still quite far from being reachable, in particular in the case
of perfect fluid sources. Therefore, research
has turned to less ambitious objectives. An obvious way
to generalize dust (i.e. stress-free) models, is to 
try to ``add" non-vanishing stresses.

A model in which analytical treatment appears feasible is that
of vanishing radial stresses, since  
in this case the general exact solution
is known in closed form~\cite{gm,gm1}
(the opposite 
case, in which only a radial stress is present,
has also been recently considered~\cite{goncalves&jhingan01a}).
The formation and nature of singularities 
in gravitational collapse with tangential stresses
has recently been studied 
quite extensively~\cite{hin98,hin,jhingan&magli00,hik}.
In the models analyzed so far, a tendency to {\em 
uncover} part of the singularity spectrum was observed, i.e., initial configurations 
that would otherwise end up in black holes develop locally naked 
singularities (in one special case, the singularity was even
shown to be timelike~\cite{hin98}, in contrast with the naked 
singularities of inhomogeneous dust collapse, which are 
typically null). 
However, all such models belong
to a very special subclass of solutions with tangential stresses, the so-called
Einstein cluster. This is a system of counter-rotating particles, in which the stress
is generated by angular momentum. 
From the physical point of view, the Einstein cluster 
is interesting 
since it mimics the effects of rotation without 
introducing deviations from spherical symmetry.
However,
although such a system 
can be formally obtained by the choice of a particular 
function of state within the general exact solution, 
this function is essentially a ``centrifugal potential"
and therefore does not fulfill the physical characteristics 
which are typical of the state equations of 
matter continua, like those to be expected
in strongly collapsed matter states (e.g. in neutron stars).
As a consequence, the results obtained for the Einstein cluster
cannot be straightforwardly extended to the general case 
of tangential stresses, which is of course interesting in its own right.

Thus motivated,  
in the present paper we carry out a complete analysis
of the final state of gravitational collapse with tangential stresses
for (marginally bound) realistic matter configurations. 
We show that the final state depends
on the first non-vanishing term of the Taylor expansion 
of the initial density distribution near the center (as 
happens to be the case in dust models) and on the first non-vanishing term of 
the Taylor expansion of another function which carries all the relevant 
information coming from the choice of the material.
It follows that the final fate of the collapse depends
on the choice of two integers, $n$ and $k$, giving the order of these 
two expansions, respectively. 
As we shall see below, there is a critical value for $k$, above 
which the tangential pressure effects are negligible, and the 
endstate of collapse is indistinguishable from that of dust models. 
Below the critical value, however, tangential stresses come into 
play, and all the configurations that would otherwise end up in a 
black hole, terminate in a singularity that is at least locally 
visible. At the critical parameter, a transition behavior is 
observed, wherein the singularity may be visible, depending on the 
details of the initial data. Both homogeneous and inhomogeneous 
density evolutions are examined in detail, and they are found to be 
qualitatively equivalent. 

This paper is organized as follows. In Sec. II, we present 
a short but self-consistent account to recall the physical 
framework and to prepare the needed mathematical formalism.
The results are given in Sec. III, where the structure of the spectrum of endstates is presented and discussed. The paper ends with concluding remarks in Sec. IV.

Geometrized units, in which $G=c=1$, are used throughout.

\section{Spherical collapse with tangential stresses}

\subsection{The general solution}

In this section we review briefly the main properties 
of spherically symmetric gravitating 
systems with vanishing radial stresses (for details, see 
\cite{gm,gm1}). 

The mathematical structure of the field equations
is simplified by the fact that the Misner-Sharp mass is conserved. 
This actually allows 
complete integration in the so called
mass-area coordinates, in which the line element 
reads
\begin{equation} ds^2 = 
-K^2\left(1-\frac{2m}{R}\right) dm^2+2 \frac{KE}{uh} dRdm - \frac 
1{u^2} dR^2 + R^2 d\Omega^{2} \ . \label{m01} 
\end{equation} 
Here 
$K$, $u$ and $h$ are functions of $R$ and $m$, and 
$d\Omega^{2}=d\theta^{2}+\sin^{2}\theta d\phi^{2}$ is the 
canonical metric of the unit two-sphere. The function $u$ is the 
modulus of the velocity of the collapsing shells and satisfies 
\begin{equation} u^2 =-1 + \frac{2m}{R} + \frac{E^2}{h^2} \ , 
\label{j} \end{equation} while \begin{equation} K = g(m) + \int 
G(m,R) dR \ , \label{m10} \end{equation} where \beq G(m,R) := 
\frac{h}{RE} \left[1+\frac{R}{2}\left(\frac{E^2}{h^2}\right)_{,m} 
\right]\left(-1 + \frac{2m}{R} + \frac{E^2}{h^2} \right)^{-3/2}. 
\label{m11} \eeq The functions $g(m)$ and $E(m)$ are arbitrary. 
The quantity $h(m,R)$ is a measure of the internal energy stored 
in the material, and can thus be regarded as an effective equation 
of state (the physical characterization of this function will be given below).
It can be shown that the energy density 
$\epsilon$ and the tangential stress $\Pi$, can be written in 
terms of $h$ as: 
\bqa 
\epsilon&=&\fr{h}{4{\pi}uKE{R^2}}, \label{dens} \\ 
\Pi&=&-\fr{R}{2h}\frac{\partial h}{\partial R}\epsilon. \label{press} 
\eqa 
These formulae show that the tangential stress 
vanishes whenever the function $h$ is independent of $R$. In such 
a case, the material is not sustained by any internal stress and 
the line element~(\ref{m01}) reduces to the one describing
spherically 
symmetric dust in mass-area coordinates \cite{ori}. Since, in all formulae, only the 
ratio of $E$ and $h$ appears, the value of $h$ along an arbitrary 
curve $R=R_0(m)$ can be rescaled to unity, so that, in 
particular, the dust solutions can be characterized by $h=1$. The 
function $R_0(m)$ plays the same role as that played by the 
initial mass distribution in the standard comoving coordinates 
[the inverse transformation from mass label to comoving label 
being $r=R_0(m)$], while the function $E(m)$ is 
the specific binding energy function.

The above recalled structure shows that a solution with tangential 
stresses is identified---modulo gauge transformations---by a 
triplet of functions $\{g,E,h\}$. This parameterization contains 
the dust spacetimes as the subset $\{g,E,1\}$. In this way, we can 
construct in a mathematically precise fashion, a comparison 
between dust and tangential stress evolutions, by comparing the 
end state of the dust collapse ($\{g,E,1\}$ with chosen $g$ and 
$E$) with the endstates of the tangential stress solutions 
$\{g,E,h\}$ with the same $g$ and $E$, and different choices of the 
equation of state $h$.

The physical singularities of the spacetime described by the 
metric (\ref{m01}) correspond to infinite energy density, and are 
given by $R=0$ or $K = 0$. At $R=0$, the shells of matter collapse 
to zero proper area, thereby leading to shell-focusing 
singularities, whereas the vanishing of $K$ implies intersection 
of different shells of matter, corresponding to shell-crossing 
singularities.
If occurring, such singularities are 
gravitationally weak  and we shall not deal with 
them in the present paper.

Non-central 
shells which become singular ($R=0$ and $m\neq0$) are always 
spacelike---thus covered---in spherical spacetimes with radial 
stresses, as a consequence of mass conservation~\cite{gm1}; 
only the central singularity $R=m=0$ can be visible, since 
at $R=m=0$, the apparent horizon and the singularity form 
simultaneously. The visibility is 
determined by the existence (or lack thereof) of outgoing null 
geodesics with past endpoint at the singularity. 
To investigate the existence of such geodesics,
we use the general method originally developed by Joshi 
and Dwivedi~\cite{joshi&dwivedi91}. One defines the quantity 
$x=R/2m^{\alpha}$ (where $1/3<\alpha\leq 1$) and observes
that, at the singularity, both $R$ and $m$ vanish. Using l'H\^ opital's
rule, one can arrive to an equation which, in the present model,
takes the form 
\beq 
x_0=\lim_{\stackrel{m\rightarrow0}{\scriptscriptstyle R\rightarrow0}} 
\frac{m^{\frac{3}{2}(1-\alpha)}}{2\alpha} \left[ -\frac{ R_{0,m} 
h(m,R_0)}{E(m) u(m,R_0)} + \int^{2m^{\alpha}x}_{R_0}G dR \right] 
\sqrt{(-1+\frac{E^2}{h^2}) m^{\alpha-1}+\frac{1}{x}} 
\left(\frac{E}{h} -\sqrt{-1+ \frac{E^2}{h^2} + 
\frac{m^{1-\alpha}}{x}} \right). \label{root0} 
\eeq 
If this 
equation admits real positive-definite roots, i.e. 
at least one finite, positive solution $x_0$ exists
for some $\alpha$, then this root represents the tangent to
an outgoing null geodesic meeting the singularity in the past, which is
therefore at least locally naked (covered otherwise). 

\subsection{Equation of state, regularity, and physical reasonability conditions} 

As recalled in the previous section, the choice of the matter model in a solution
with tangential stresses corresponds to the choice of the function $h$. This is
equivalent to the choice of an equation of state like, e.g., the barotropic
equation of state connecting density and pressure in the case of 
a perfect fluid.
The choice of $h$ is thus restricted by physical considerations, as follows 
\cite{jkgm}.

The internal energy per unit volume of the material ($\epsilon$, say)
is given by $\epsilon=\rho_b h$, 
where $\rho_b$ is the baryon number density. If the material
is physically viable, the internal energy has to coincide with the number density
in a (locally) relaxed state of the continuum. This obviously means, as in the 
classical (i.e. non-relativistic) mechanics of continua, 
that $h$ must be a positive, convex function having a minimum in the locally relaxed state.
Without loss of generality, we take the value of the minimum equal to one. Since all deformations 
are described ``gravitationally" in the frame we are working with, the state of local relaxation
corresponds to the flat space values of the metric components. It follows
that the most general physically valid equation of state has the form 
\beq
h(m,R)=1+\g(m)(R_{0}-R)^{2} + h_2(m,R), \label{eos1} 
\eeq 
where $h_2(m,R)$ goes to zero more rapidly than $(R_0-R)^2$, as $R$ tends to $R_0$.
The function $\g(m)$ is positive and models the ``strength'' 
of the tangential pressure. With the scaling $R_{0}(m)=r$, the 
squared factor on the right-hand-side lies in the interval 
$[0,r^{2}]$, where the lower and upper limits are realized on the 
initial and singular slices, respectively. Since we are interested only in the behavior near the central singularity, 
it follows that, as far as the causal structure of the singularity is concerned, 
we can consider the function $h$ 
to be of the form
\beq
h(m,R)=1+\beta_{k}m^{k/3}(R_{0}-R)^{2} , \label{eos}
\eeq
where $k$ must be positive in order to assure 
regularity at the center, and 
the positive constant $\beta_{k}$ measures the effects of tangential 
stresses. The case of dust collapse corresponds to 
$\beta_{k}=0$, while for $\beta_k$ non-zero the 
first non vanishing term of the Taylor expansion 
of the tangential stress near the center is of order $k$.

Having chosen a suitable matter model, 
to address the issue of cosmic censorship, 
one must also ensure that several other reasonability conditions are satisfied.

First of all, the spacetime must have a regular center (so that the singularity
forming at $r=0$ is a true dynamical singularity).
This requires 
the functions $ K$, $u$, and $E/h$ to 
be smooth and bounded away from zero. 
Further, we require the collapse 
to proceed from a regular initial slice, where the 
metric and the exterior curvature are continuous, the physical quantities 
are bounded, and 
there are no trapped surfaces. 
As we have seen, initial data consists of two independent quantities
$g$ and $E$, which can be equivalently
described by the initial mass [$m(r)$, i.e., $R_0(m)$] 
and the initial velocity distribution 
$f:=E^2-1$. In what follows, we take $f=0$, 
i.e., we consider marginally bound configurations.
Regularity on the initial surface 
implies that 
\beq 
m(r)=F_{0}r^{3}+F_{n}r^{n+3}+{\mathcal O}(r^{n+4}), \label{mss} 
\eeq where $F_{n}$ denotes ($4\pi/3$ times) the first 
non-vanishing derivative of the initial energy density profile at 
the origin. Positivity of mass requires 
$F_{0}>0$, and $F_{n}<0$ for any realistic configuration, where 
the energy density decreases away from the center. 

We want 
to exclude the presence of trapped surfaces on the initial slice. This 
requires $2m<R$ on the $R_{0}(m)=r$ surface: \beq 
\left(\fr{2m}{R}\right)_{R=r}=\fr{2m(r)}{r}=2F_{0}r^{2}+{\mathcal 
O}(r^{n+2})<1, \eeq where the second equality holds only near the 
origin. The first equality imposes a maximum size on the matter 
distribution, for a given density profile. Near the origin, the 
requirement $2m/R<1$ is always satisfied on the initial slice, for 
$0\leq r<1/\sqrt{2F_{0}}$. 

Finally, and in accord with the spirit of the cosmic 
censorship conjecture~\cite{penrose69}, we require that 
the matter content obeys the weak energy condition (WEC): 
\beq 
\epsilon>0\;, \; \epsilon+\Pi>0. \eeq 
From Eq. (\ref{dens}), one 
sees that the first inequality is satisfied provided $h>0$ and 
$K<0$ (recall that $u<0$, for the collapsing situation we are 
interested in). From Eq. (\ref{eos}), it follows that 
$\partial_{R}(\ln h)<0$, and hence the tangential stress is positive, 
which is a sufficient condition for the second inequality (positivity of $\Pi$ implies that
also the strong energy condition, $\epsilon+2\Pi \geq 0$, is automatically satisfied here). 

\subsection{The final state of collapse} 

Dust models do not admit globally 
regular solutions, i.e., singularity-free solutions. This is not 
necessarily true for 
general matter fields, even within spherical symmetry. Examples of 
globally regular solutions in spherical symmetry include 
certain classes of perfect fluid models, which evolve 
to an eternally oscillating configuration from regular initial 
data (see e.g. ~\cite{bonnor&faulkes67}), and solitonic solutions of the 
massive Einstein-Klein-Gordon system~\cite{seidel&suen91}. 
Accordingly, it is important to understand the qualitative 
behavior of the dynamics, from a given set of regular initial 
data, before addressing the causal nature of the final 
state. 

The equation of motion~(\ref{j}) for the collapsing shells can be 
re-written as \begin{equation} 
u^2=\frac{Z(R,R_0)}{R[1+\g(m)(R_0-R)^2]} \ , \label{motion} 
\end{equation} where the ``effective potential'' $Z$ is defined by 
\begin{equation} Z(R,R_0):=2m+\g(m)(2m -R)(R_0-R)^2 \ . \label{pot} 
\end{equation} This function may be regarded as the analogue of 
the Newtonian effective potential governing the motion of the {\em 
fixed} shell $R_0$, wherein the allowed regions of the motion 
correspond to $Z\geq0$. Setting $R=\zeta(t, R_0) R_0$, to the lowest order 
in $R_0$ we obtain 
\beq 
2 F_0[1+\beta_km^{k/3}(R_0-R)^2] \geq \beta_k 
m^{k/3}\zeta(t).
\eeq 
This condition always holds, independently of the 
choice of $k$, as $m$ goes to zero.
Therefore, there is always a region of initial data leading
to continued gravitational collapse.

We now would like to check whether the central point ($m=0$) 
eventually gets trapped. Since $m$ is conserved for any given 
shell, this can be done by analyzing the dynamics of shells near 
the center, along curves $R=\lambda m$ with $\lambda 
>2$~\cite{hik}.
The requirement $Z\geq0$ now reads \beq 2 \geq \beta_k 
m^{k/3} (R_0-\lambda m)^2(\lambda -2) \ . \eeq Clearly, near the 
center this inequality always holds, which implies that 
non-central shells inevitably become singular. This is in contrast 
with the Einstein cluster case, 
where rotation has the effect that all the shells near the 
central one remain regular for all times~\cite{hin}. 
In the present case, instead, the apparent horizon inevitably forms, 
and therefore the central singularity may or may not be visible, 
depending on the existence of geodesics meeting this singularity
in the past.

\section{The spectrum of endstates}

\subsection{The root equation} 

The analysis of the root equation for gravitational collapse
with tangential stresses is non-trivial, 
because of the integral appearing on 
the right-hand-side of Eq. (\ref{root}), which introduces a 
``non-local" behavior. Indeed, as we have seen,
only the case of the Einstein cluster 
has been analyzed so far.

To check whether the root
equation has positive-definite solutions, we observe that, 
although the integral appearing in square brackets cannot be carried
out in terms of elementary functions, its behavior near the
center can be analyzed as follows.
First, let us denote 
\begin{equation} 
I_{1}(m;\alpha):=\int_{R_0}^{2m^{\alpha}x} G(m,R)dR, \label{test} 
\end{equation} where $G(m,R)$ is defined by Eq. (\ref{m11}). Due to 
the mean value theorem, there exists $\chi(m) \in 
(R_0,2m^{\alpha}x)$ such that 
\begin{equation} I_{1}(m;\alpha):= 
(2m^{\alpha}x -R_0) G(m,\chi(m)). \label{funda} 
\end{equation} 
Since $R_0 \sim m^{1/3}$ as $m$ goes to zero, $1/3< \alpha \leq 
1$, and $x$ is positive-definite, it follows that both 
$m^{-1/3}\chi (m)$ and $m^\alpha/\chi (m)$ are also 
positive-definite as $m\rightarrow0$. One can thus evaluate the 
right hand side of Eq. (\ref{funda})---using Eq. (\ref{m11})---as 
follows: 
\bqa 
&&I_{1}(m;\alpha) = 
\nonumber 
\\ 
&&
m^{-1} \frac{(2m^{(3\alpha-1)/3}x 
- F_0^{-1/3} +\cdots )}{2\sqrt{2}} 
\left(\frac{\chi}{m^{1/3}}\right)^{1/2} 
\left[1-\frac{\chi}{m^{1/3}}\left(\frac{\g[F_0^{-1/3} 
-(\chi/m^{1/3})]^2}{2[1+\g(R_0-\chi)^2]}\right) \right]^{-3/2} 
\nonumber 
\\ 
&&
\left[1-\frac{\chi}{m^{1/3}}\left(\frac{(m\g_{,m})[F_0^{-1/3} 
-(\chi/m^{1/3})]}{2[1+\g(R_0-\chi)^2]^2}\right) 
-\frac{\chi}{m^{1/3}} 
\frac{(\g F_0^{-1/3})[F_0^{-1/3}
-(\chi/m^{1/3})]}{[1+\g(R_0-\chi)^2]^2}\right] 
\nonumber 
\\ 
&&
[1+ 
\g(R_0-\chi)]^{1/2}, \label{needed} \eqa 
where the dots stand for 
terms of higher order in positive powers of $m^{n/3}$. Since 
$\g(0)=0$, the divergence is driven by ${\mathcal O}(m^{-1})$, with 
all the other terms remaining finite.
Therefore, the limit \begin{equation} I_{2}(m;\alpha):= 
\lim_{m\rightarrow0} \int_{R_0}^{2m^{\alpha}x} m \ G(m,R)dR , 
\label{life} \end{equation} is convergent. 
Using Lebesgue's 
dominated convergence theorem, we can expand the integrand near 
the center ($m = 0$) in leading powers of $m$, and integrate 
successive terms~\cite{details}. 
It follows that
the root equation, 
for a general density distribution [given by Eq. 
(\ref{mss})], with the equation of state (\ref{eos}),
can be written as:

\beq 
x_0=\lim_{\stackrel{m\rightarrow0}{\scriptscriptstyle 
R\rightarrow0}} 
\frac{m^{\frac{3}{2}(1-\alpha)}}{\alpha} 
\left(\Theta_n m^{-1+n/3} 
+B_k m^{-1+k/3} + 
\frac{2m^{3(\alpha-1)/2}}{3} x^{3/2}+ \dots \right) 
\left(\frac{1}{\sqrt{x}} -\frac{m^{(1-\alpha)/ 2}}{x} \right), 
\label{root} \eeq 
where

\beq
\Theta_n:=-\frac{n F_n}{18{\sqrt 2} F_0^{(2n+9)/6}},
\eeq
\beq
B_k :=
\frac{{\sqrt {2}}(9+k)\beta_k}{945 F_0^{3/2}}.
\eeq

\subsection{The dust limit}

In order to better visualize the effects of tangential stresses 
in spherical dust collapse, it is important to recall briefly
how to recover the (very well known) dust
case from our general set-up of the previous section
(for a complete review and list of references see \cite{review}).
This is easily  
done by setting $\g(m)=0$, and thus $\beta_{k}=0$. 
The root equation becomes then 

\beq 
x_0=\lim_{m\rightarrow0} \frac 1{\alpha}\left[\Theta_{n} 
m^{(3+2n-9\alpha)/6}+\fr{x^{3/2}}{3}+\dots\right] 
x^{-1/2}\left[1-x^{-1/2}m^{(1-\alpha)/2}\right]. \label{rltb} 
\eeq 
It 
is straightforward to check that for 
$\alpha\neq\alpha_{n}\equiv\fr{1}{3}(1+\fr{2n}{3})$ the root 
equation does not admit any positive-definite solution: if 
$\alpha<\alpha_{n}$, then $x_{0}=0$, and for $\alpha>\alpha_{n}$ 
the right-hand-side of Eq. (\ref{rltb}) diverges in the 
$m\rightarrow0$ limit. One must therefore have $
\alpha=\alpha_n$. In this case, 
the 
equation reduces to
\beq 
x_{0}^{3/2}=
\fr{1}{\alpha}
\left(\Theta_{n}+\fr{x_{0}^{3/2}}{3}\right)
\left[1-x_{0}^{-1/2}\lim_{m\rightarrow0} m^{(3-n)/9}\right]. 
\eeq 
Clearly, we must have $n\leq3$, for $x_{0}\in(0,+\infty)$. 
For $n<3$, we obtain \beq 
x_{0}=\left[(1+\fr{3}{2n})\Theta_{n}\right]^{2/3}, \eeq and the 
singularity is locally visible.
For $n=3$, after a little algebra, one obtains the quartic 
equation (where $Z\equiv \sqrt{x_{0}}$): \beq 
2Z^{4}+Z^{3}+\gamma (1-Z)=0, \eeq 
with
$$
\gamma=\Theta_{3}=F_{3}F_{0}^{-15/6}/(2\sqrt{2}) \ .
$$ 
Standard results from polynomial theory can be used to show that 
this quartic has real and positive roots iff 
$\gamma>\gamma_{c}$, where~\cite{review} 
\beq
\gamma_c:=
(26+15\sqrt{3})/4. \label{xic}
\eeq

\subsection{Structure of the endstates}

For $\beta_k \neq 0$, 
the root equation is given by Eq. (\ref{root}). 
It is immediately seen that  one can 
always choose $1 \leq n \leq 3$ 
{\it and/or} $1 \leq k \leq 3$, in such a way that, 
for suitable $1/3<\alpha\leq1$, this 
equation gives a finite, algebraic condition.
It then follows that in all such cases the singularity can be naked.

For $n<3$ {\it or} $k<3$ the singularity is naked,
since a positive root always exists
(we omit here tedious, but straightforward calculations;
the values of the roots are reported in Table I). 

For $k\geq4$, the effects of 
tangential stresses are negligible, and the endstate is exactly 
the same as that in dust collapse, with the shape of the initial 
central density profile determining the existence of 
positive-definite roots, and thus local visibility. The very same can be said,
however, about the effects of inhomogeneity for $n\geq 4$, since  
{\em all} such
configurations would lead to black holes, while now they all terminate 
in a visible singularity if $k<3$. 

A transitional (``critical") behavior is observed
at the three entries 
forming the ``boundary" of the $n<3$, $k<3$ region.
Here, naked singularities occur iff 
a quartic equation has positive solution(s).
This quartic 
is identical in form to that holding for dust, namely
\beq 
2Z^4 +Z^3 
+\gamma_{nk} (1-Z) =0, \label{quar} 
\eeq 
where $\gamma_{nk}$ takes the values 
\bqa 
\gamma_{33}&=&\gamma +\delta, \\ 
\gamma_{34}&=&\gamma, \\
\gamma_{43}&=&\delta, 
\eqa 
with
\beq
\delta :=\frac{4\sqrt{2} \beta_3}{105F_0^{3/2}}. 
\eeq
Therefore, at each ``boundary", 
naked singularities are formed 
iff the corresponding quantity 
$\gamma_{nk}$ 
is greater than the quantity $\gamma_c$ defined in (\ref{xic}).

\begin{table}[h] \centering \begin{tabular}{ccccc} 
Initial data & n=1 & n=2 & n=3 & n $\geq$ 4  \\ \hline
k=1    & $\frac{-21F_1+8\beta_1 F_0^{1/3}}{84{\sqrt 2}F_0^{11/6}}$  & $\frac{{\sqrt
2}\beta_1}{21 F_0^{3/2}}$ & $\frac{{\sqrt 2}\beta_1}{21 F_0^{3/2}}$ &
$\frac{{\sqrt 2}\beta_1}{21 F_0^{3/2}}$  \\ 
k=2    & $\frac{-F_1}{4{\sqrt 2}F_0^{11/6}}$ &
$\frac{-105F_2+22\beta_2 F_0^{2/3}}{420{\sqrt
2}F_0^{13/6}}$ & $\frac{11\beta_2}{210{\sqrt 2} F_0^{3/2}}$ &
$\frac{11\beta_2}{210{\sqrt 2}
F_0^{3/2}}$  \\ 
k=3   & $\frac{-F_1}{4{\sqrt 2}F_0^{11/6}}$ & $\frac{-F_2}{4{\sqrt 2}F_0^{13/6}}$ &
$\gamma_{33}>\gamma_{c}$  & $\gamma_{43}>\gamma_{c}$  \\ 
k$\geq$4 & $\frac{-F_1}{4{\sqrt 2}F_0^{11/6}}$ & 
$\frac{-F_2}{4{\sqrt 2}F_0^{13/6}}$ & $\gamma_{34}>\gamma_{c}$  & no positive real roots  \\ 
\end{tabular} 
\caption{The endstate of collapse for inhomogeneous 
collapse with tangential pressures. See text for details.}
\end{table}  

\subsection{The homogeneous case} 

The homogeneous dust solution is the well-known
Oppenheimer-Snyder spacetime~\cite{os} (the seminal paradigm for black hole formation, for over sixty years), whose dynamical evolution results in a Schwarzschild black hole, with an interior, spacelike (hence covered) singularity.
To evaluate the root equation for this case, we 
simply set $F_{n}=0$ in Eq. (\ref{rltb}), obtaining: 
\beq 
x_{0}^{1/2}=\lim_{m\rightarrow0} 
\fr{1}{3\alpha}\left[x^{1/2}-m^{(1-\alpha)/2}\right]. 
\eeq 
Clearly, we need $\alpha\leq1$, or else the right-hand-side of the 
above equation diverges negatively in the $m\rightarrow0$ limit. 
For $\alpha=1$, we obtain $x_{0}^{1/2}(2x_{0}^{1/2}+1)=0$, which 
has no positive-definite solution, and the singularity is 
therefore covered. For $\alpha<1$, the $x_{0}$ terms drop out, and 
a self-consistent solution exists iff $\alpha=1/3$, which is {\em 
not} allowed. Hence, the singularity is always covered, as 
expected. 

This example provides a particularly simple test-bed 
for studying the effects of tangential pressure on the final state of 
collapse. 
In presence 
of tangential stresses, from Eq. (\ref{root}) with $\Theta_n=0$, we obtain: 
\beq 
x_0=\lim_{\stackrel{m\rightarrow0}{\scriptscriptstyle 
R\rightarrow0}} 
\frac{1}{\alpha} 
\left[B_k
m^{(3+2k-9\alpha)/6} +\frac{x^{3/2}}3+ \dots 
\right]\left[\frac{1}{\sqrt{x}} -\frac{m^{(1-\alpha)/2}}{x} 
\right]. \label{rt} 
\eeq 
A curious phenomenon occurs here. In fact, 
the above equation is formally identical to the root equation
for {\it inhomogeneous} dust [cf. Eq. (\ref{rltb}) above]
with $k$ and $B_k$ playing the role of $n$ and $\theta_n$ respectively.
Clearly, in the leading term we can 
always choose $1 \leq k \leq 3$, such that $1/3<\alpha\leq1$, and 
the singularity may thus be visible. 
Again, there is a clear 
transition at $k=3$, below which (stronger tangential stresses) 
the singularity is always visible, and above which (weaker 
tangential stresses) is always covered.
At the transitional value, $k=3$, we have again 
to solve a quartic of the kind (\ref{quar}) with 
parameter equal to $\delta$, so that naked singularities occur if
$\delta>\gamma_c$. The final state of collapse is 
summarized on Table II, below. 

\begin{table}[h] \centering \begin{tabular}{ccc} 
Initial data  &  Root $(x_0)^{3/2}$     & Singularity  \\ \hline
$k=1$    & $\displaystyle{\frac{{\sqrt 2}\beta_1}{21 F_0^{3/2}}}$ & visible \\ 
$k=2$    & $\displaystyle{\frac{11\beta_2}{210{\sqrt 2} F_0^{3/2}}}$ & visible \\  
$k=3$     & $\delta>\gamma_{c}$ & transition \\ 
$k\geq4$  & no real positive roots                      & covered \\
\end{tabular} 
\caption{The endstate of homogeneous 
collapse with tangential pressure.}\end{table}

\section{Concluding Remarks} 

We have examined here the effects of tangential stresses on 
the final state of (marginally bound) collapse. 
Since the nature 
of the singularity turns out to depend only on the Taylor expansion of the data 
and of the state function near the center, 
it is possible to classify the final state of all the solutions 
that have a physically reasonable equation of state. 
This classification can be done
in terms of the ``strength'' of the tangential stress near 
the center. If this strength is above a certain threshold, the 
tangential pressure dominated collapse always leads to a locally naked singularity, 
independently of the choice of the 
initial data. If, instead, the tangential pressure is sufficiently weak, 
then the final state is the same as that occurring 
in dust collapse, where 
the shape of the initial density profile near the origin fully 
determines the visibility of the singularity. 

These results actually disproof a conjecture,
recently put forward by one of us~\cite{gm1}, the proposal of which,
roughly speaking, was that the space
of the data leading to naked singularities in spherical 
collapse could not be ``larger" than that of dust initial data
leading to the same singularities. Contrary to this expectation, the present results are consistent with quite a different scenario: they 
support the notion that the strong version of the cosmic censorship 
conjecture is likely to be violated for realistic matter distributions
in a way that can safely be described as {\it common}, insofar as the matter content is concerned. Indeed, 
our analysis shows that 
naked singularities are {\em not} an artifact of dust models,
or a byproduct of ``toy" models 
constructed without giving the equation 
of state of the collapsing matter, or, finally, 
a spurious outcome of unnatural effects induced by rotation in spherical symmetry.

There remains the possibility that such naked singularities are an artifact of spherical symmetry by itself, although both numerical
and analytical results suggest that dynamical curvature singularities may persist in non-spherical collapse~\cite{shapteuk,wang&pereira00,alvi&goncalves&jhingan01}.

\section*{Acknowledgments} The authors thank K. Thorne and H. Kodama for useful discussions and/or comments. SMCVG 
acknowledges the support of FCT (Portugal) Grant PRAXIS 
XXI-BPD-163301-98, and NSF Grants AST-9731698 and PHY-0099568. SJ 
acknowledges the support of Grant-in-Aid for JSPS Fellows No. 
00273, and the hospitality of the California Institute of 
Technology, where part of this work was done.


\begin{thebibliography}{99}
\bibitem{details} 
The proof is analogue to that given for the Einstein cluster in the Appendix of Ref.~\cite{jhingan&magli00} above.
\end{thebibliography}
\end{document}